# Stripping Efficiency and Lifetime of Carbon Foils

W. Chou[a,*], M. Kostin[b] and Z. Tang[a]

[a]*Fermilab, P.O. Box 500, Batavia, Illinois 60510, U.S.A.*

[b]*NSCL, 1 Cyclotron, East Lansing, Michigan 48824, U.S.A.*

*Abstract*

Charge-exchange injection by means of carbon foils is a widely used method in accelerators. This paper discusses two critical issues concerning the use of carbon foils: efficiency and lifetime. An energy scaling of stripping efficiency was suggested and compared with measurements. Several factors that determine the foil lifetime – energy deposition, heating, stress and buckling – were studied by using the simulation codes MARS and ANSYS.

[*] Corresponding author. Tel: +1-630-840-5489; fax: +1-630-840-6039; E-mail: chou@fnal.gov

## 1. Introduction

Like many other laboratories, Fermilab employs a charge-exchange method during the injection of particle beams from the Linac to the Booster. The $H^-$ ions are accelerated to 400 MeV in the Linac and pass through a thin carbon foil when entering the Booster. The foil strips two electrons from each ion and converts the ions from $H^-$ to $H^+$, which are then accelerated to 8 GeV in the Booster. Figure 1 shows the foil changer in the Booster and Figure 2 are used carbon foils. In the future, Fermilab plans to replace the Linac and Booster by a superconducting rf linac, nicknamed “Proton Driver” [1]. This machine will accelerate $H^-$ ions up to 8 GeV with total power of 0.5 MW. These $H^-$ ions will then be stripped to protons in foils and injected into the 120 GeV Main Injector for mass production of neutrinos aimed at a detector (MINOS) in mine shaft in Soudan, Minnesota to study neutrino oscillations.

When the energy of $H^-$ ions increases, it becomes more difficult to convert $H^-$ to $H^+$ in foils because the interaction cross sections are decreased at higher energies. One obvious way to compensate for the cross section reduction is to use thicker foils. Unfortunately this approach has limitations since it leads to more severe foil heating and stress, which would reduce the foil lifetime. This problem is especially important for high intensity hadron accelerators, in which minimal beam loss and proper foil lifetime are essential to machine operations.

In this paper, we introduce an energy scaling of the stripping efficiency of carbon foils. It is based on two known measurements at 200 MeV and 800 MeV, respectively. The energy deposition and heating are studied numerically by the code MARS, whereas the mechanical stress and buckling by another code ANSYS. Throughout the paper, Fermilab accelerators will be used in numerical examples. The same analysis can be applied to other machines, such as J-PARC (Japan) and SNS (U.S.A.).

## 2. Stripping efficiency of carbon foils

### *2.1. Theory on cross section*

The theoretical approach for calculating the collisional electron-detachment cross section for negative hydrogen ions incident on hydrogen, helium, oxygen, nitrogen and other gas targets can be found in numerous publications. Here we will use the results from Gillespie in Refs. [2-5].

Gillespie's method is an extension of Bethe's theory. It employs the sum-rule technique in the Born approximation to sum over all excited final states of the $H^-$ ion for calculating the total electron loss cross section. This method is particularly useful in our case because $H^-$ ion has no bound excited states. The total electron loss cross section can be expressed as:

$$(\sigma_{-1,0} + \sigma_{-1,1}) = 8\pi a_0^2 \left(\frac{\alpha^2}{\beta^2}\right) \sum_{n\neq 0} \sum_{m} \left[I_{nm} - J_{nm}(\beta^2) - K_{nm}(\beta^2)\right] \qquad (1)$$

in which $\sigma_{-1,0}$ is the cross section from $H^-$ to $H^0$, $\sigma_{-1,1}$ from $H^-$ to $H^+$, $n$ the final states of $H^-$, $m$ the final states of the target atom, $a_0$ the Bohr radius, $\alpha$ the fine structure constant, $\beta$ the relativistic factor, $I_{nm}$, $J_{nm}$ and $K_{nm}$ integrals. The first integral $I_{nm}$ is the asymptotic (high-energy) leading order contribution to the cross section and is independent of the incident velocity. The second and third integrals ($J_{nm}$ and $K_{nm}$) are the next order correction terms for low energies. By performing these integrals explicitly, Gillespie was able to obtain numerical results for the case of $H^-$ ions incident on various target atoms and claimed to be in agreement with experimental data, as shown in Figures 3 and 4 [2,3].

*2.2. Energy scaling of stripping efficiency of carbon foils*

It is interesting to note that the physics governing the foil stripping and residual gas stripping is the same. It is only because of the enormous difference in atom density between foil and residual gas that the $H^-$ ions can travel thousands of meters in the transport line free of stripping and, suddenly, be fully stripped by a foil in a distance of a few μm!

When $H^-$ ion energy increases, the cross section decreases as shown in Section 2.1. As a consequence, the stripping efficiency also decreases. This is a major concern for high energy (e.g. 8 GeV) $H^-$ injection, because low efficiency implies high injection losses.

Several cross section measurements of $H^-$ ion incident on carbon foil at different energies have been reported [6-11]. For example, Figure 5 shows the data at 200 MeV measured by Webber and Hojvat [8] and Figure 6 the data measured by Gulley et al. at 800 MeV [7].

The stripping efficiency at the two energies is remarkably different. For instance, when a 200 μg/cm$^2$ foil is used, only 0.4% $H^0$ remains after the foil at 200 MeV, whereas the number of $H^0$ is increased to 11.2% at 800 MeV. In order to estimate the stripping efficiency at other energies, we invoke the energy scaling of the cross section in Eq. (1), namely, the cross section decreases asymptotically as $1/\beta^2$, where $\beta$ is the relativistic factor of $H^-$ ions.

We based our scaling on the 800 MeV data, because they have higher accuracy. We first scaled the 800 MeV data to 200 MeV and compared them with the measured data at 200 MeV. We found them in good agreement and this gave us confidence. We then scaled the 800 MeV data to 400 MeV and 8 GeV. The predicted 400 MeV cross sections will be compared with the planned measurement at the Fermilab Booster. Table 1 is a summary of these numbers.

Figure 7 is a plot of the $H^0$ population for the 5 cases listed in Table 1 at different foil thicknesses. The two curves for 200 MeV, one measured and another calculated using the scaling, lay almost completely on top of each other. This indicates the scaling works well in this energy region. Because the scaling is asymptotically correct at high energies, it is expected to work even better at 8 GeV.

Based on this estimation, the carbon foil thickness is chosen to be 600 $\mu g/cm^2$ for the 8 GeV Proton Driver, which corresponds to 0.5% $H^0$ population in this analytic model. Alternatively, one may employ two consecutive foils with a thickness of 300 $\mu g/cm^2$ of each.

## 3. Lifetime of carbon foils

### *3.1. Energy deposition*

We will use Fermilab Proton Driver as an example in this section. It has two injection scenarios: 90-turn injection that has a beam current of 25 mA and pulse duration of 1 ms, 270-turn injection that has a beam current of 8 mA and pulse duration of 3 ms. In each case the total number of particles is the same, namely, $25 \times 10^{-6}$ Coulomb, or $1.56 \times 10^{14}$ $H^-$ ions. There are three particles in each $H^-$ ion: one proton and two electrons. Therefore the number of particles is about $N = 4.7 \times 10^{14}$.

The energy loss of moderately relativistic particles other than electrons in matter is primarily due to ionization and atomic excitation. For 8 GeV protons incident on a carbon foil, the stopping power $|dE/dz| = 1.847$ MeV/(g/cm$^2$) [12]. When an electron travels together with an 8 GeV proton, its kinetic energy is 4.357 MeV. The stopping power, is 1.71 MeV/(g/cm$^2$) [13].

In the present design, there are two foils separated by a 40 cm gap. The dimension of the foils is 12 mm × 12 mm. Each foil has a thickness of 300 μg/cm$^2$. The energy deposition on each foil would be 554 eV by one proton and 513 eV by one electron. While the electrons would hit the foil only once, the protons would hit it multiple times during injection. Simulation shows the average number of hits for each proton is 4.4 (90-turn) and 15.9 (270-turn), respectively. Taking these into account, the total energy deposition on each foil during injection is, respectively, 0.0608 J (by protons, 90-turn injection), 0.2199 J (by protons, 270-turn injection), and 0.0256 J (by electrons). The injection interval is 1.5 sec.

The beam size is 0.8 × 0.4 cm$^2$, and the distribution is Gaussian. Using $\sigma_x = 0.2$ cm and $\sigma_y = 0.1$ cm, the energy deposition per unit area $D(x,y)$ can be written as:

$$D(x,y) = \frac{A}{2\pi\sigma_x\sigma_y} \exp\left(\frac{-x^2}{2\sigma_x^2}\right) \exp\left(\frac{-y^2}{2\sigma_y^2}\right) \tag{2}$$

where $A$ is the total energy deposition listed above. The space distributions of particles for each turn were simulated by the code STRUCT [14]. The results are then fed into the code MARS [15] where interactions of the protons and electrons in the carbon foils were simulated and the deposited energy calculated. It was assumed that the energy deposition is instantaneous and there is no evolution of the foil temperature during injection (see

Section 3.3 below). The specific heat of carbon foils is 0.165 cal/g-K or 0.6908 J/g-K at room temperature and is treated as a function of temperature in the calculation because it rises in a hot foil. The emissivity is assumed to be 0.8.

*3.2. Thermal analysis using MARS*

In the calculation, we make a conservative approximation. For the first foil, it is assumed that all $H^-$ ions are stripped in the very upstream part of the foil and electrons pass through it contributing to the heating. In the meantime, it is also assumed that a maximum of 20% of $H^-$ could survive the first foil and get stripped in the second foil.

Table 2 lists the energy deposition and temperature rise due to heating by protons and electrons for each injection scheme. It can be seen that the instantaneous temperature rise in both foils for the 270-turn injection scheme would bring the temperature close to the melting point of carbon, which is about 3600 °C. The peak energy deposited due to protons only for the 270-turn injection scheme is the same as the combined energy due to protons and electrons because of a specific space distribution of the electrons.

Figures 8-9 are graphical representations of the temperature rise after one cycle. The size of the histograms corresponds to the size of the foil.

*3.3. Thermal and mechanical analysis using ANSYS*

The thermal process is governed by diffusion. In the carbon foil, it propagates at 0.174 $cm^2/s$. During 1 ms (3 ms) beam pulse, the diffusion length is 0.0132 cm (0.0228 cm), much smaller than the foil size. The mechanical process propagates with speed of sound, which is 2558 m/s in carbon. During 1 ms (3 ms) beam pulse, the mechanical disturbance propagates 255.8 cm (767.4 cm), much larger than the foil size. Therefore, as far as the thermal process is concerned, energy deposition can be considered instantaneous. But it is static as far as the mechanical process is concerned.

A finite element model of ANSYS is built to simulate the problem. The model is supported on top and right sides (both thermally and mechanically). The interaction between the two foils (heating each other by radiation) is neglected.

In thermal analysis, both initial and boundary conditions are 275 K. Energy deposition is input as heat rate (energy divided by time). Since it occurs on a time scale which is very small compared to heat diffusion, energy deposition is instantaneous. The temperature increase of the carbon foil can be calculated just as the integration of energy deposition over density and specific heat. After the pulse, in a period of 1.5 second there will be no energy deposition. Heat will then be taken out by thermal radiation. Figure 10 is a typical temperature history at the hottest spot. The temperature cycle reaches equilibrium quickly (in two or three cycles). Maximum temperature (just after the beam pulse) is plotted in Figures 11-12.

Since the mechanical process can be considered static, the only load is the thermal stress induced in the foil. Using the maximum temperature from thermal analysis, the static displacement and stress are calculated. These results are summarized in Table 3 and plotted in Figs. 13 through 16. (Note: There is some difference in maximum temperature between Table 2 and 3. This is attributed to the different algorithm used in the codes MARS and ANSYS.)

Buckling analysis was performed on a model representing one quarter of the foil, using large deformation option in finite element. To break the symmetry, small load was applied perpendicularly at the center of the foil. The thermally induced load was then applied, and the buckled deformation was obtained after removing the small symmetry breaking load. The results are shown in Figs. 17-19. The maximum displacement is 0.038 cm. At first glance this may seem small, but in relative terms it is rather significant. Assuming a 0.0003 cm thick foil (i.e. 600 μg/cm$^2$ carbon foil of density 2 g/cm$^3$) this displacement is more than 100 times as large as the foil thickness.

*3.4. Carbon foil lifetime*

There are a number of factors that have impact on the lifetime of carbon foils: instant temperature rise, average temperature rise, mechanical stress and displacement, fatigue due to thermal buckling, sublimation (solid to gas transition at temperatures above 1600 °C), radiation damage of the structure [16], etc. Although we know how to estimate these effects either analytically or numerically, it is not clear which one is the determining factor. It is quite likely that the failure of a carbon foil is a combinational result of all these factors. Furthermore, foil manufacture technique and foil microstructure play a major role in lifetime. For the same ion bombardment, different types of foils can have vastly different lifetime. Hence, beam test in an accelerator is the ultimate way to determine the lifetime of a carbon foil.

## 4. Discussion

Liouville's theorem precludes multi-turn injection of particles identical to those already present in a circulating beam. $H^-$ injection through a stripping foil provides a mean to circumvent this difficulty. As the energy increases, however, stripping efficiencies tend to decrease, which leads to the employment of thick foils. But the thickness is limited by thermal and mechanical considerations and cannot arbitrarily increase to compensate for the reduction in interaction cross-section. Furthermore, a thick foil would result in more severe effects on the beam (e.g. emittance dilution due to multiple Coulomb scattering, acceptance limit due to large angle single Coulomb scattering, energy straggling, etc.) as well as on the machine (e.g. radiation activation of magnets nearby the foil). But these topics are beyond the scope of this paper.

In a high intensity proton machine, uncontrolled particle loss must be kept at a very low level in order to prevent activation of the accelerator components. A balance between the foil efficiency and foil lifetime must be studied carefully. It is necessary to design conservatively and ensure that adequate safety factors are introduced. Progress in foil technology demonstrates the possibility of new foils of much longer lifetime compared to the conventional ones [17]. This is encouraging news. Plans are under way to install these new foils in an operational accelerator for long term testing.

**Acknowledgements**

This paper is part of the Fermilab Proton Driver study during the period of 2004-2005. It was benefited from numerous stimulating and helpful discussions with a number of colleagues from Fermilab and other institutions, including H.C. Bryant, A. Drozhdin, J. Lackey, R.J. Macek, J.-F. Ostiguy, G.H. Rees and P. Yoon. A mini-workshop held in December 2004 at Fermilab was particularly helpful in reviewing many of the results presented in this paper [18].

**Table 1:** Cross Section of $H^{-}$ Incident on Carbon Foil (unit $10^{-18}$ $cm^2$)

| | 800 MeV (measured) | 200 MeV (measured) | 200 MeV (scaled) | 400 MeV (scaled) | 8 GeV (scaled) |
|---|---|---|---|---|---|
| $\sigma_{-1,\,0}$ | 0.676 ± 0.009 | 1.56 ± 0.14 | 1.49 | 0.942 | 0.484 |
| $\sigma_{0,\,1}$ | 0.264 ± 0.005 | 0.60 ± 0.10 | 0.584 | 0.368 | 0.189 |
| $\sigma_{-1,\,1}$ | 0.012 ± 0.006 | –0.08 ± 0.13 | 0.026 | 0.0167 | 0.0086 |

**Table 2:** Energy Deposition and Instantaneous Temperature Rise of Carbon Foil (MARS)

| | Peak Energy Deposit<br>Foil 1 (J/g) | Peak Temperature Rise<br>Foil 1 (K) | Peak Energy Deposit<br>Foil 2 (J/g) | Peak Temperature Rise<br>Foil 2 (K) |
|---|---|---|---|---|
| Electron | 1478 ± 2 | — | 296 ± 6 | — |
| Proton, 90 turns | 2182 ± 122 | — | 2230 ± 138 | — |
| Proton, 270 turns | 6616 ± 459 | — | 6639 ± 488 | — |
| e + p, 90 turns | 3621 ± 128 | 1991 ± 70 | 2502 ± 141 | 1470 ± 83 |
| e + p, 270 turns | 6616 ± 459 | 3358 ± 233 | 6639 ± 488 | 3368 ± 248 |

**Table 3:** Temperature Rise, Displacement and Stress of Carbon Foil (ANSYS)

| | Foil 1, 90 turn | Foil 1, 270 turn | Foil 2, 90 turn | Foil 2, 270 turn |
|---|---|---|---|---|
| Temperature Max (K) | 2084 | 3011 | 1675 | 2985 |
| Displacement $u_x$ (mm) | -0.0218<br>0.0090 | -0.0501<br>0.0235 | -0.0168<br>0.0078 | -0.0498<br>0.0224 |
| Displacement $u_y$ (mm) | -0.0263<br>0.0067 | -0.0838<br>0.0193 | -0.0243<br>0.0063 | -0.0839<br>0.0192 |
| Stress $\sigma_x$ (N/cm$^2$) | -7331<br>6145 | -7744<br>3873 | -5329<br>2679 | -6965<br>4142 |
| Stress $\sigma_y$ (N/cm$^2$) | -4390<br>2887 | -12418<br>3219 | -4010<br>2397 | -11896<br>4916 |

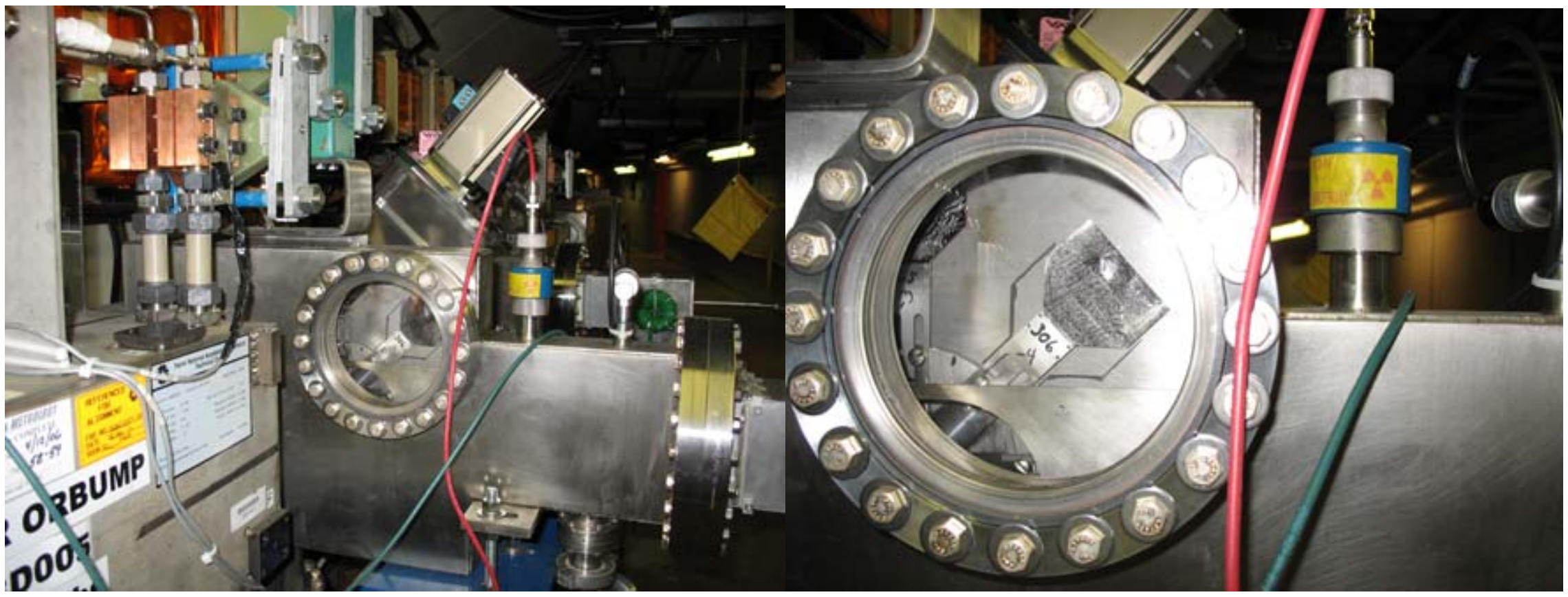

**Figure 1**

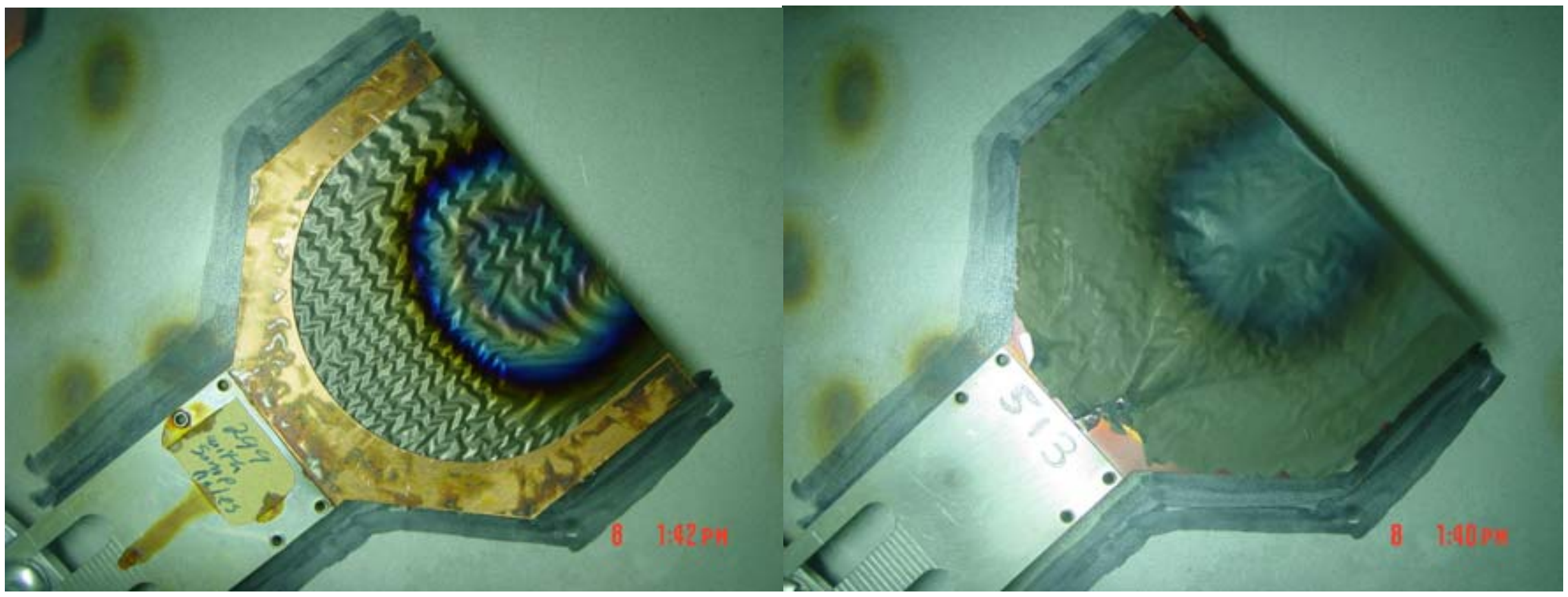

**Figure 2**

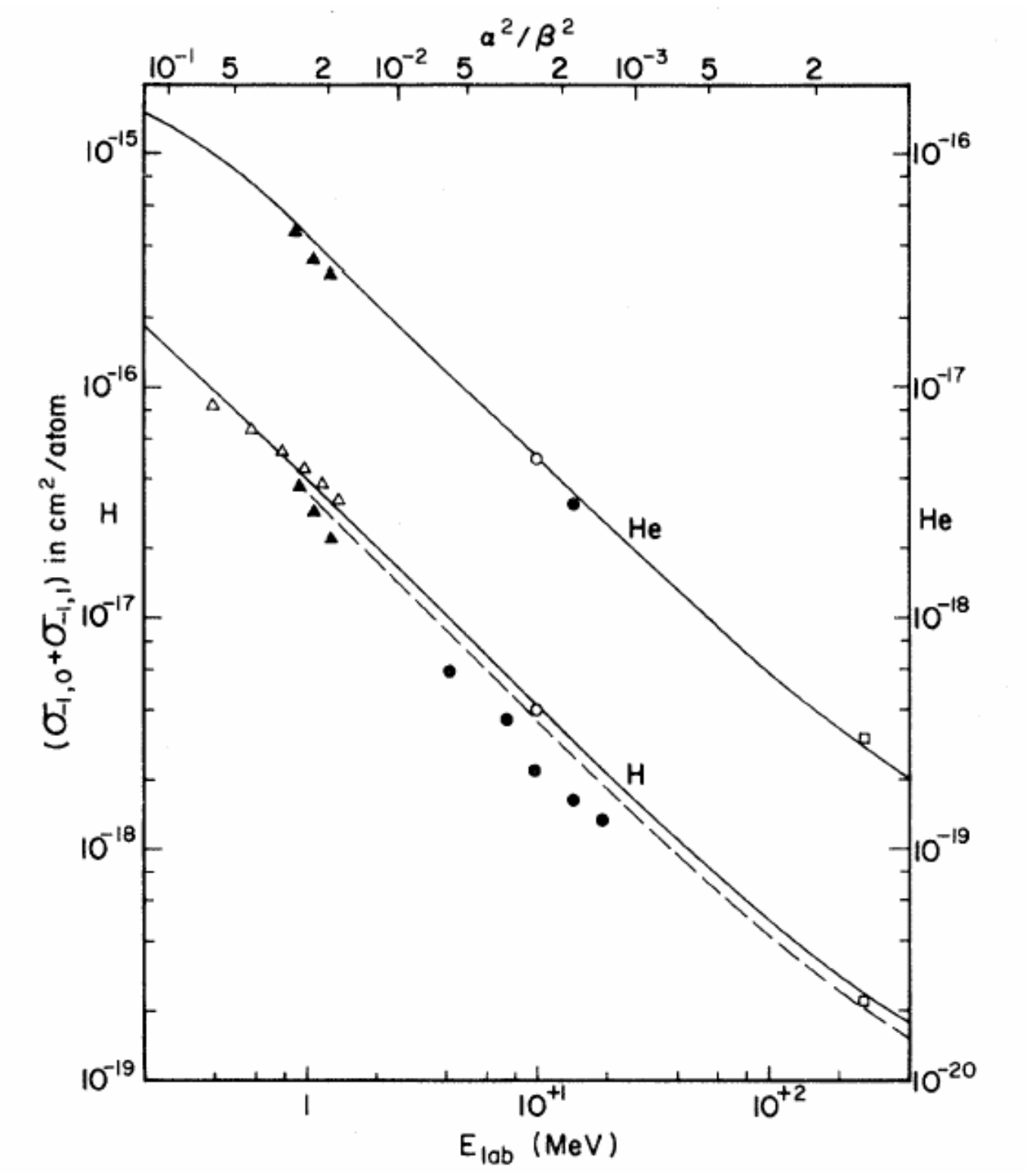

**Figure 3**

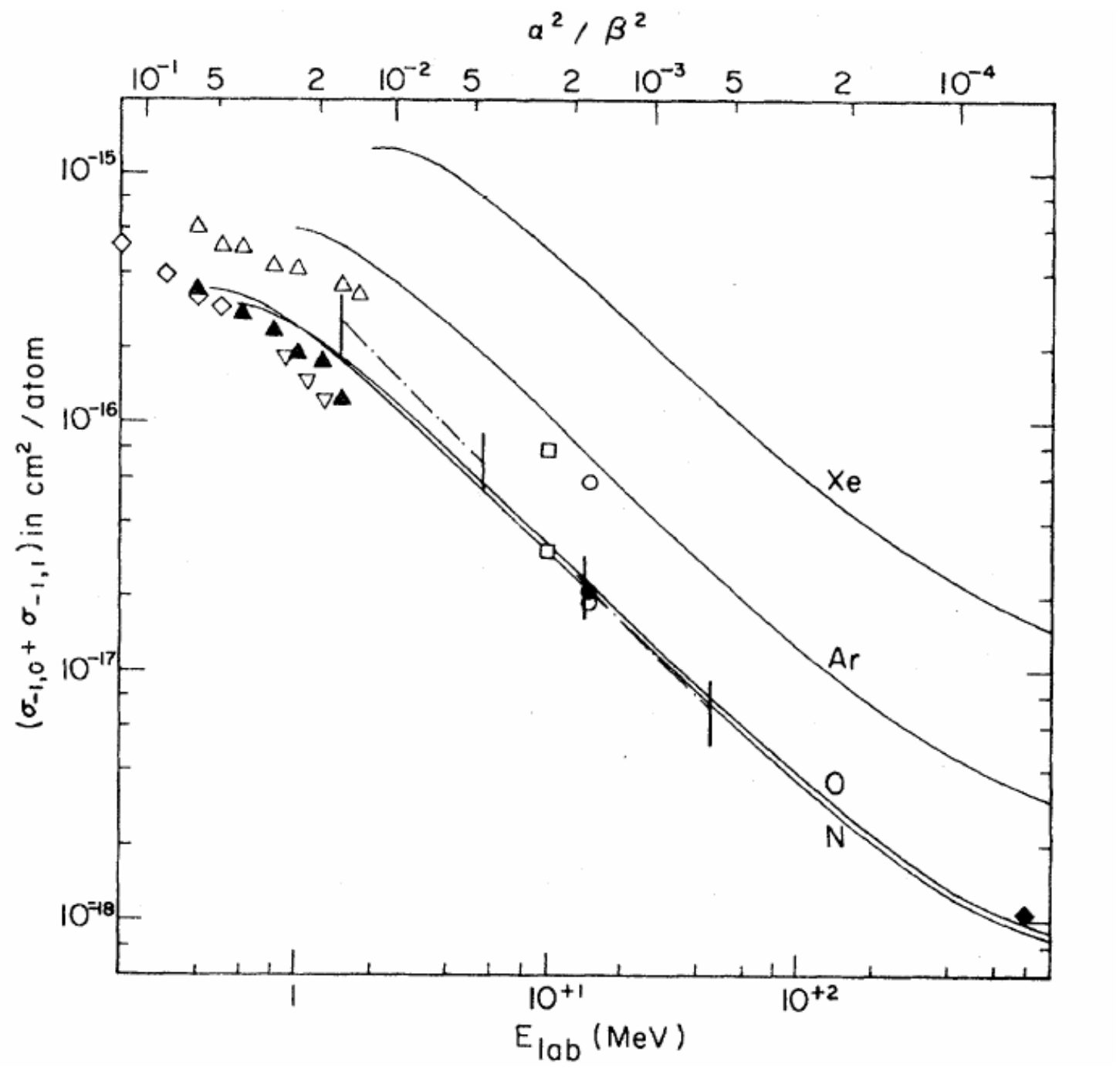

**Figure 4**

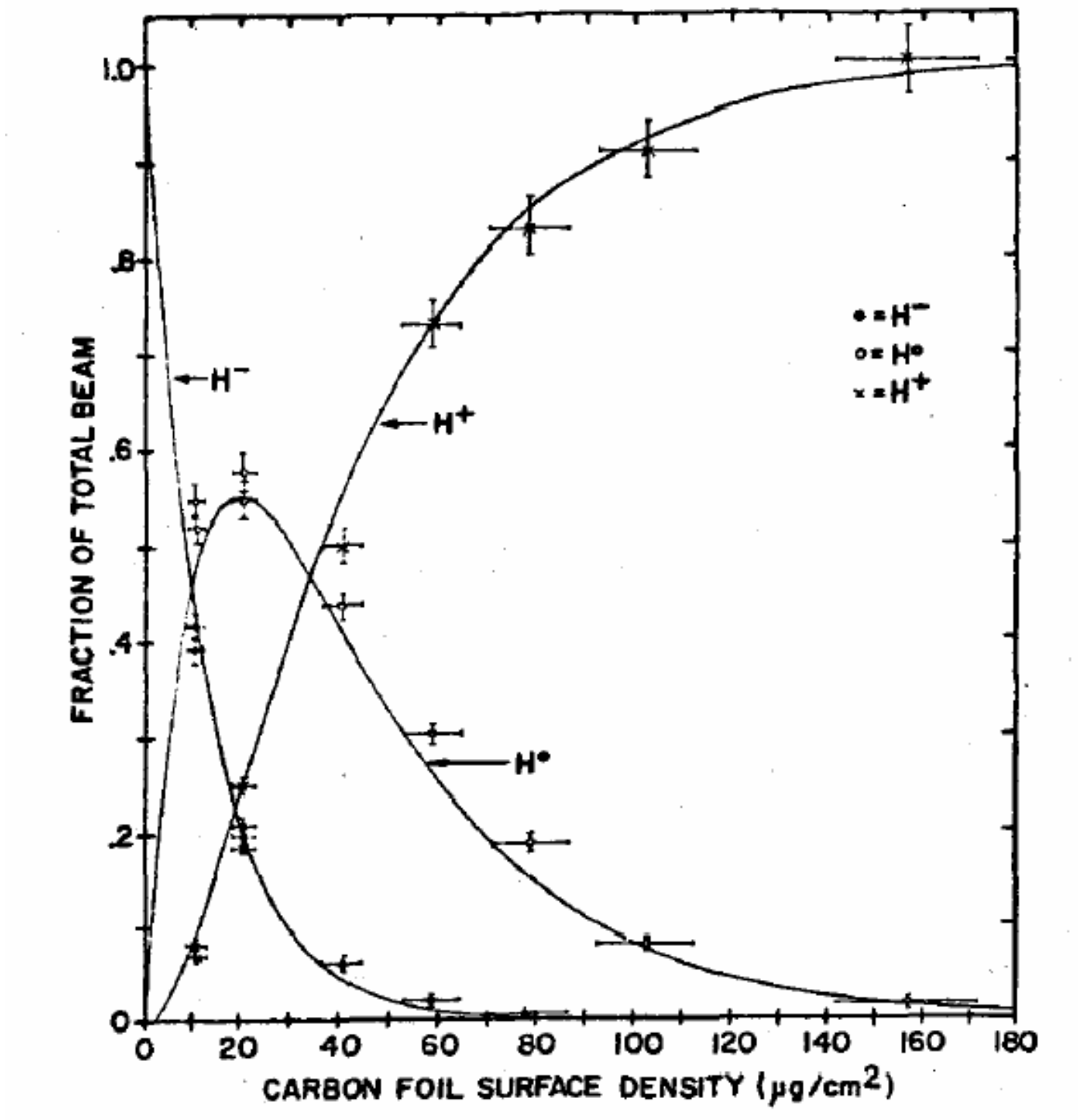

**Figure 5**

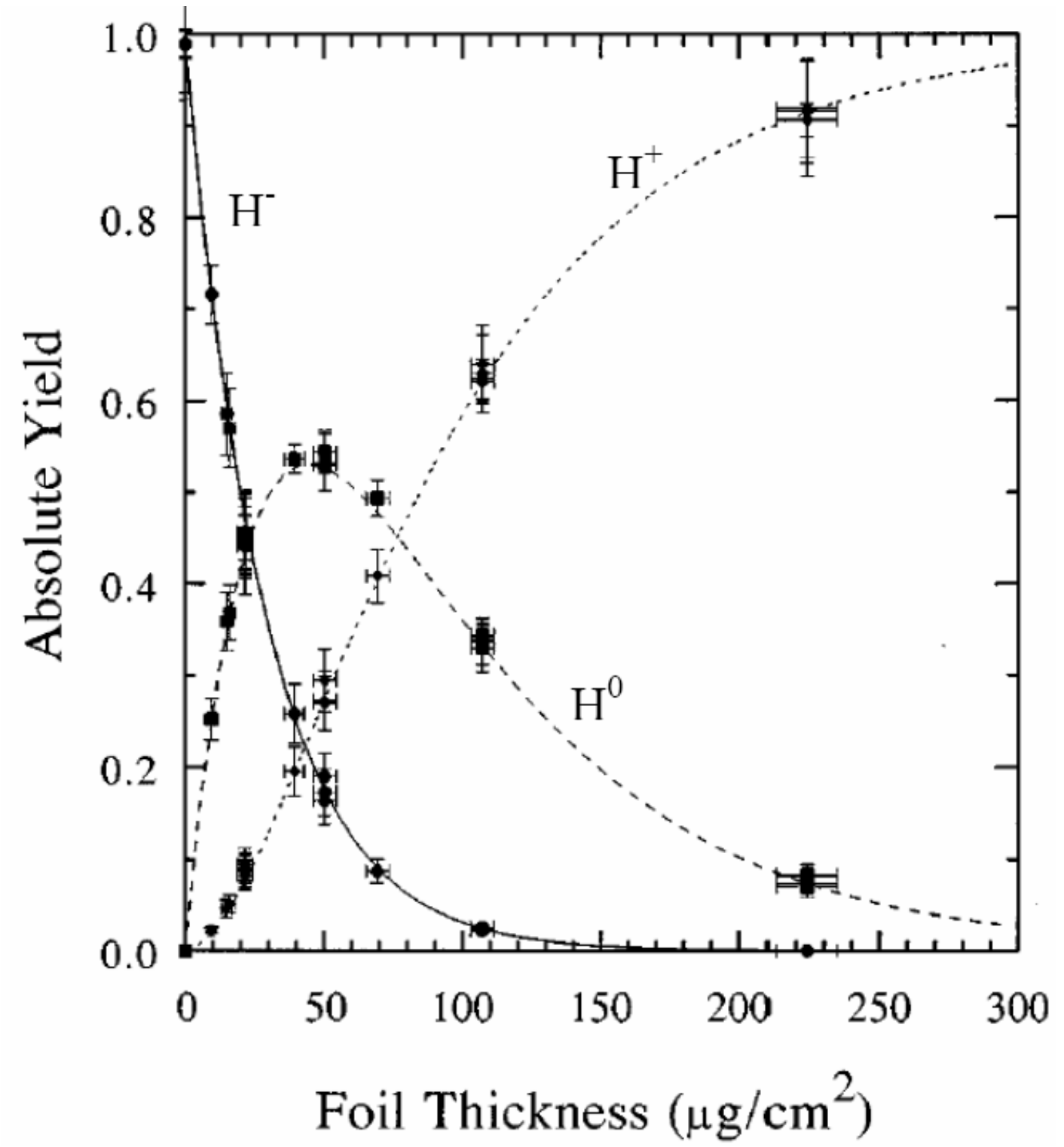

**Figure 6**

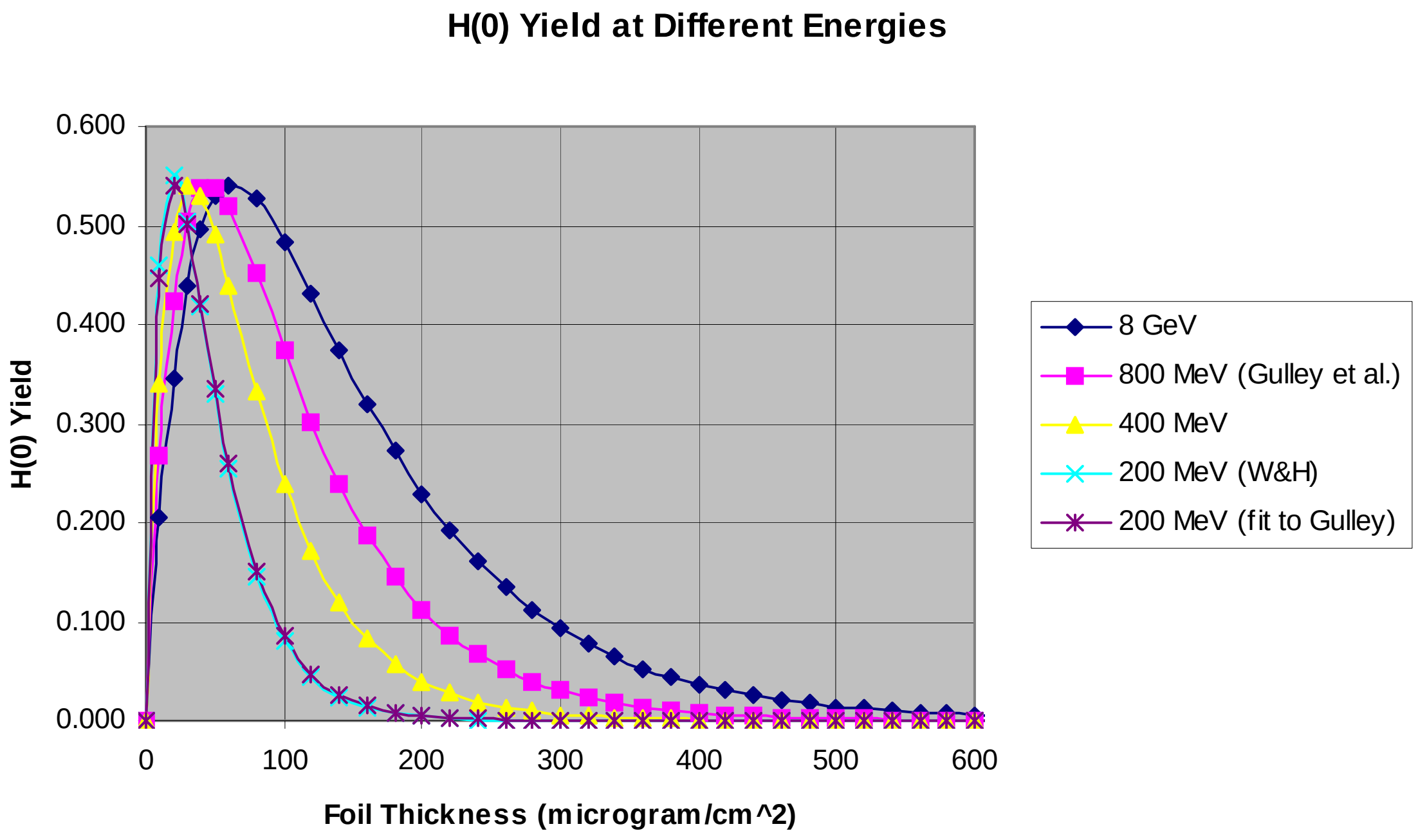

**Figure 7**

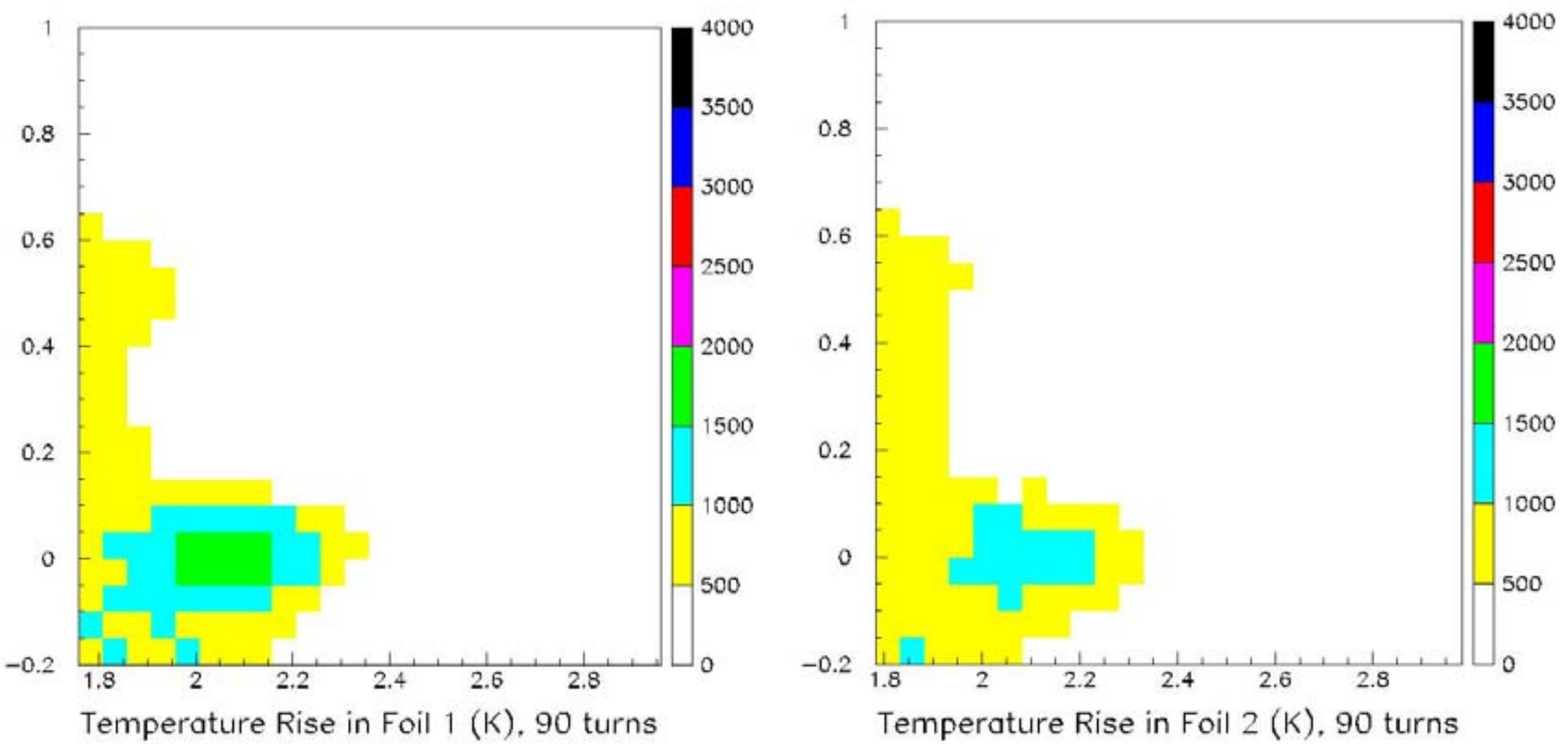

**Figure 8**

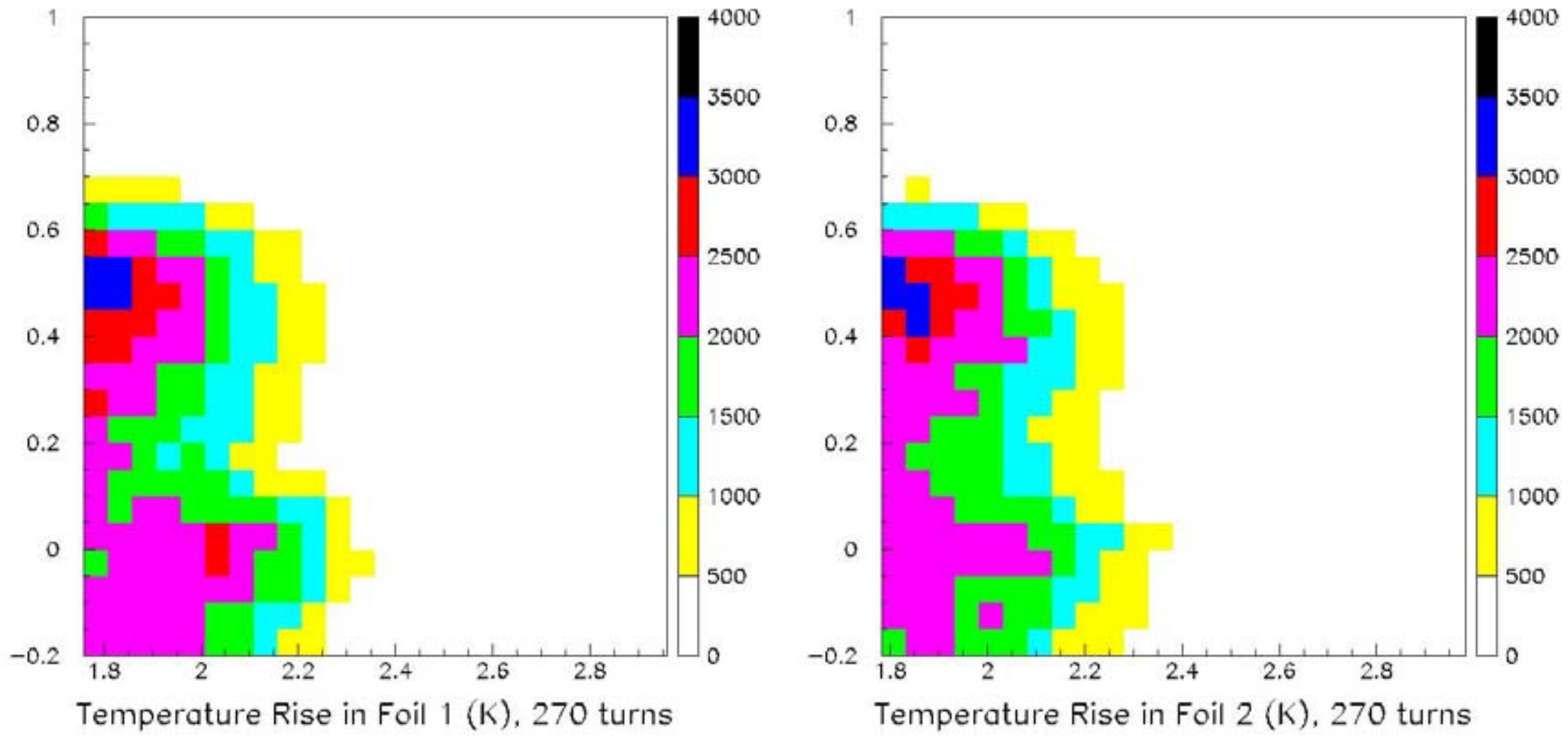

**Figure 9**

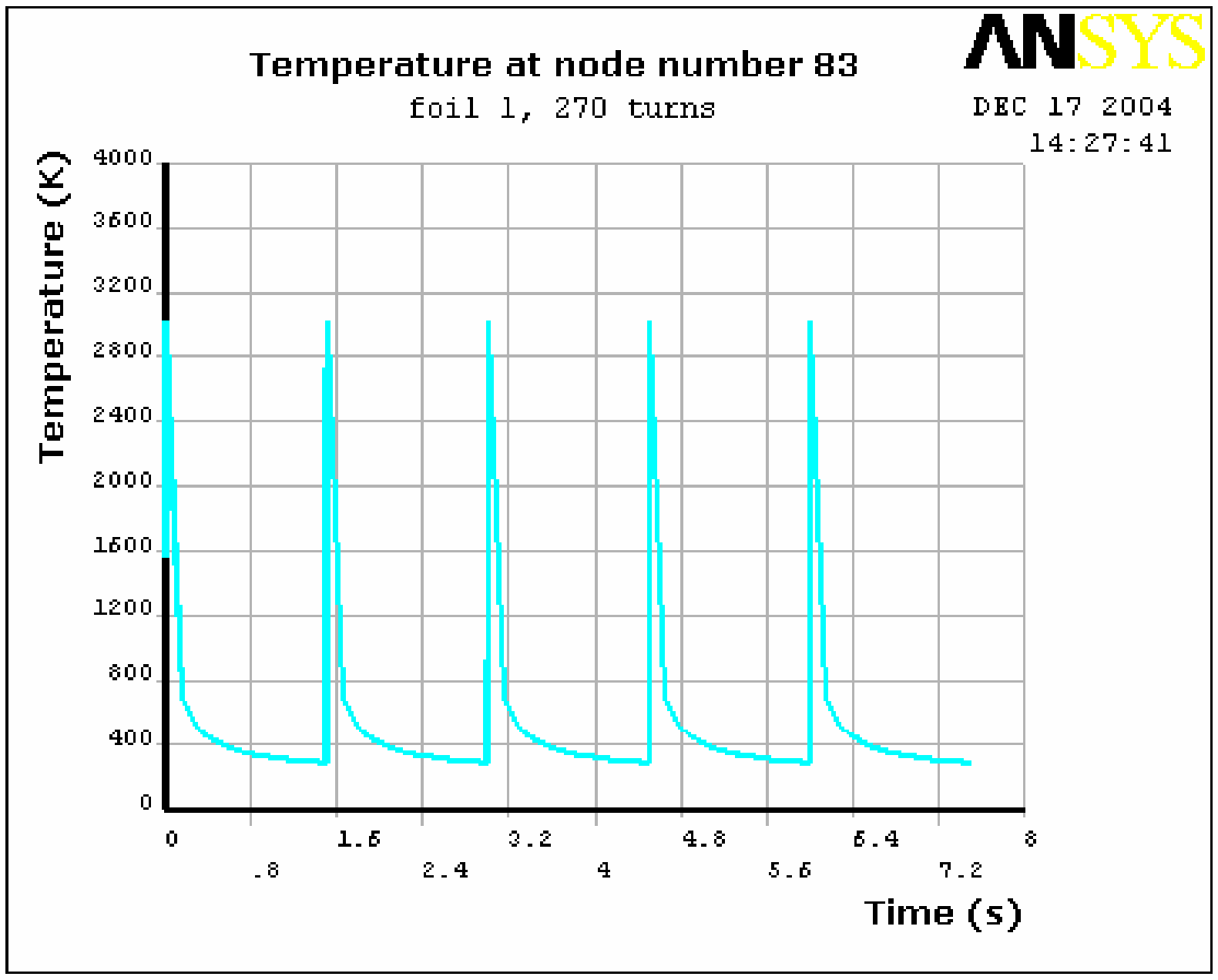

**Figure 10**

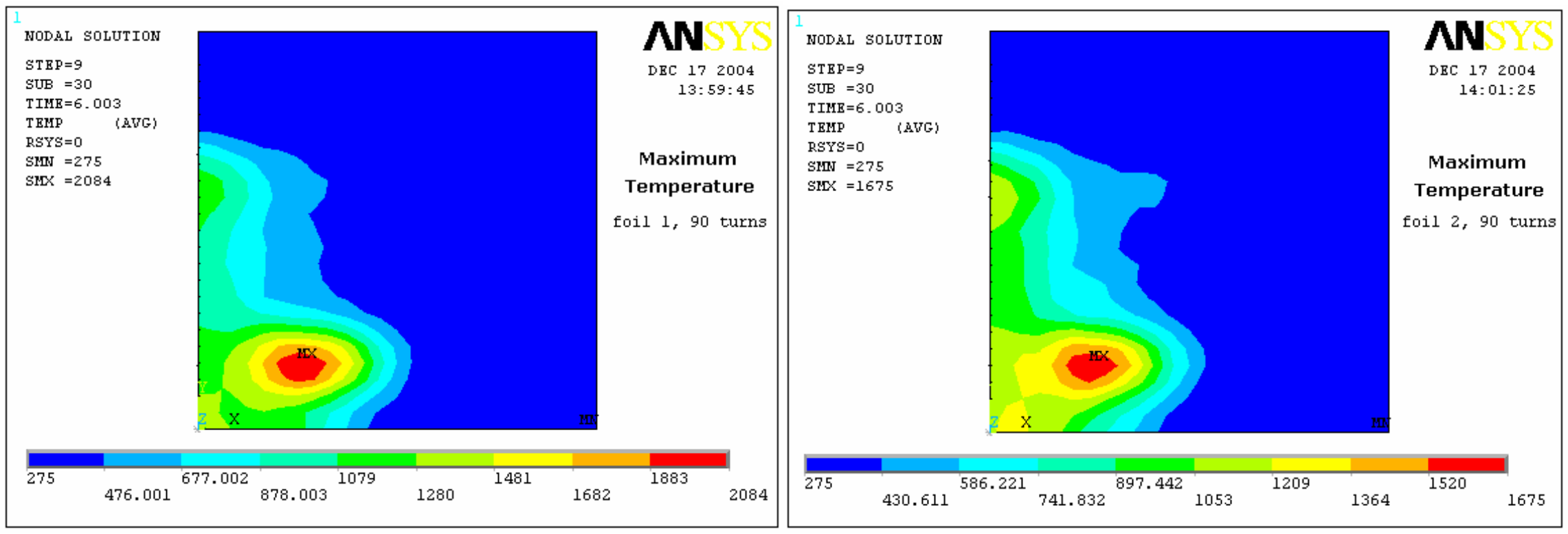

**Figure 11**

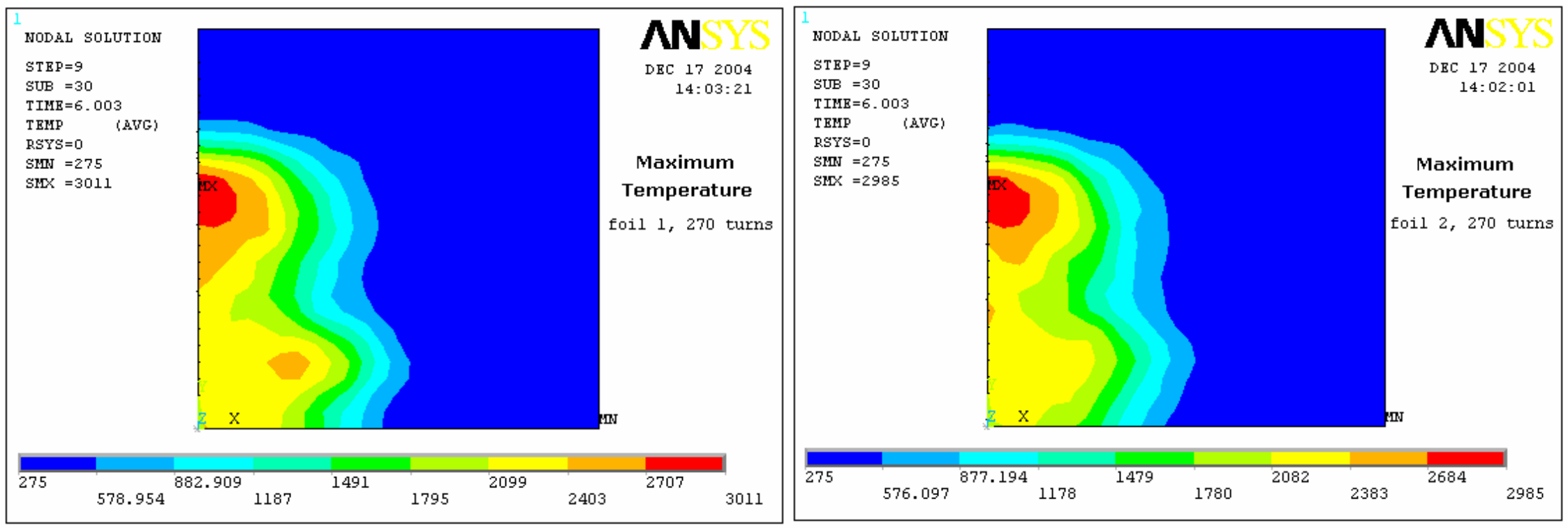

**Figure 12**

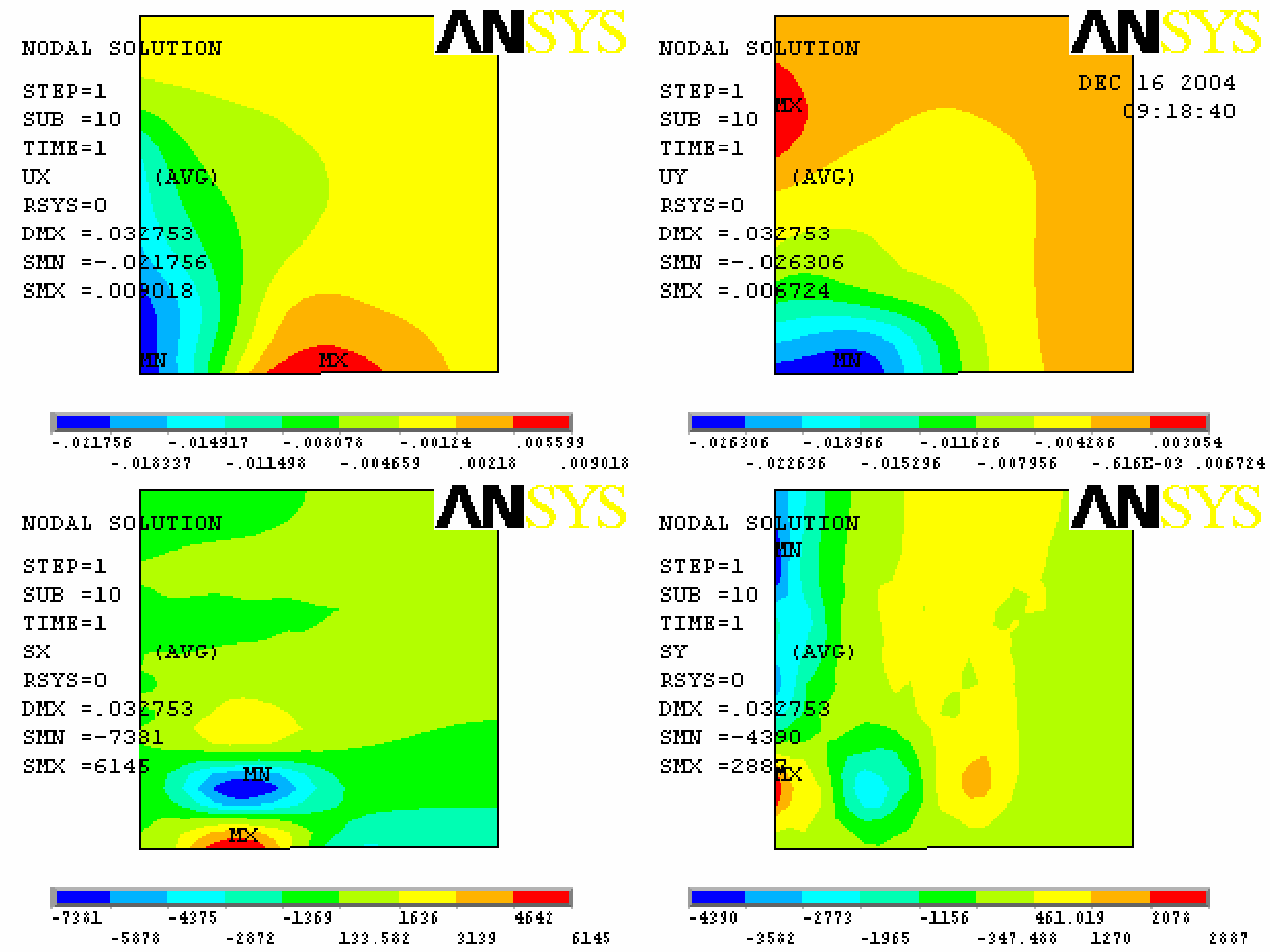

Displacement and stress (foil 1, 90 turns)

**Figure 13**

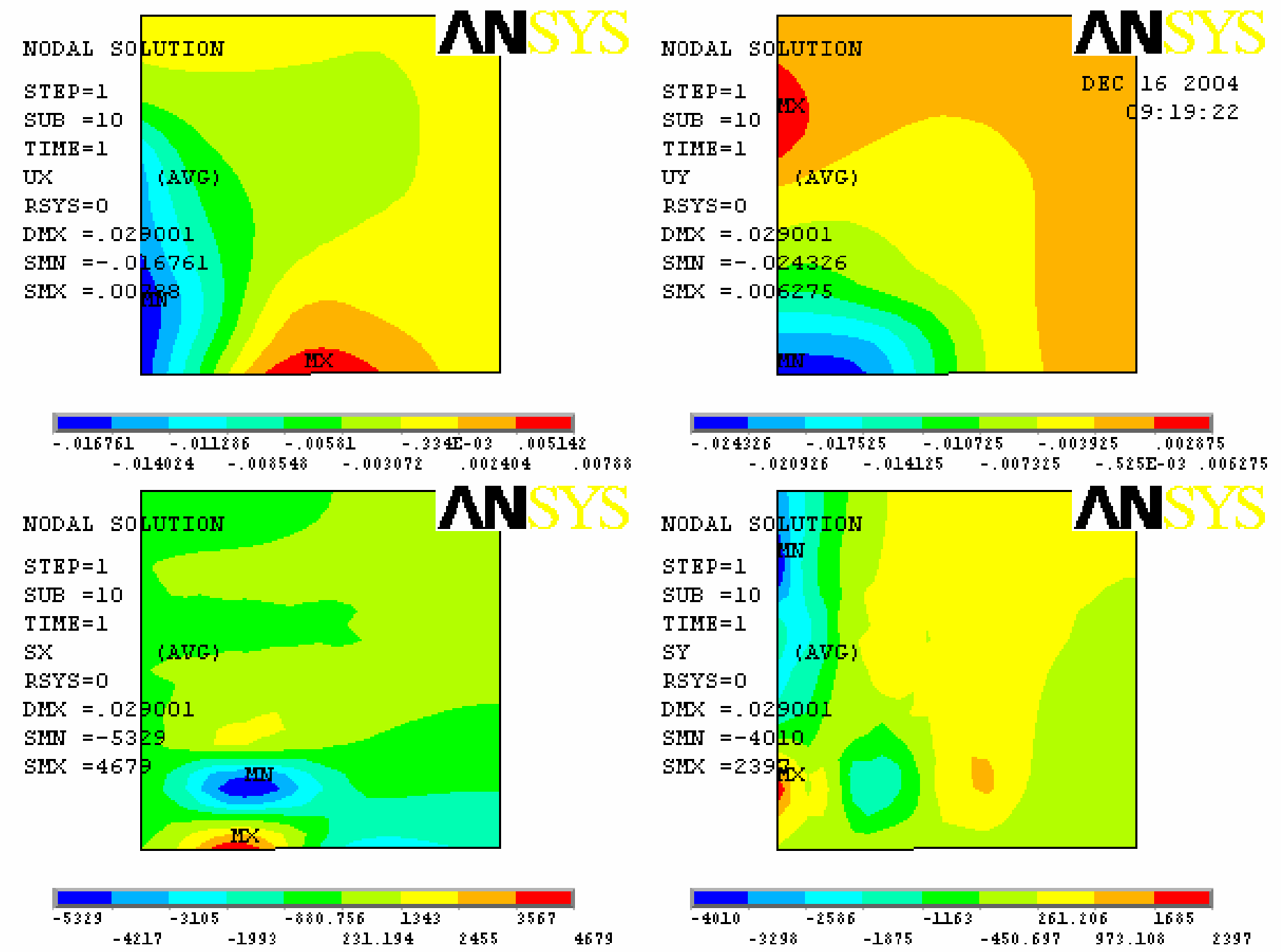

Displacement and stress (foil 2, 90 turns)

**Figure 14**

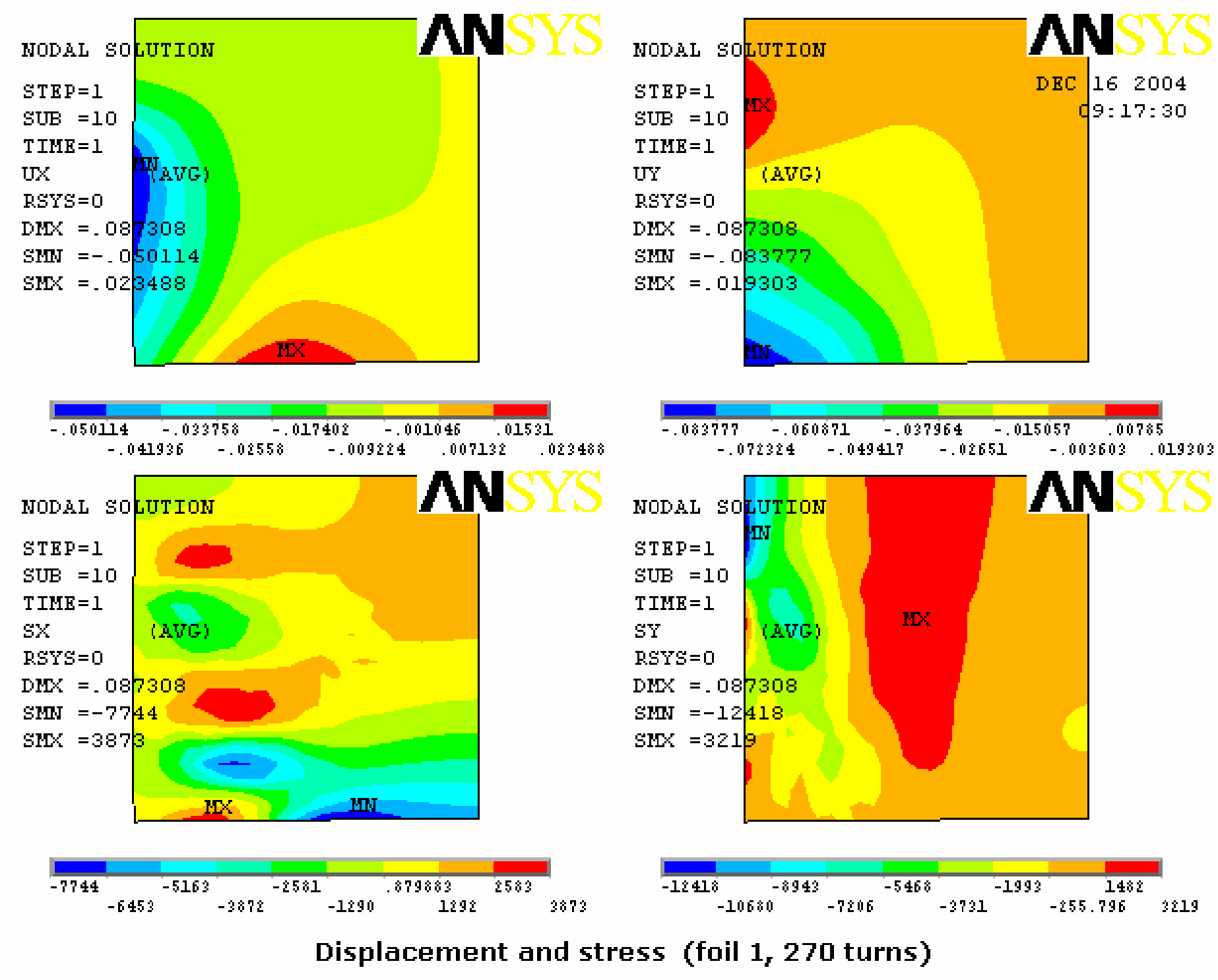

**Figure 15**

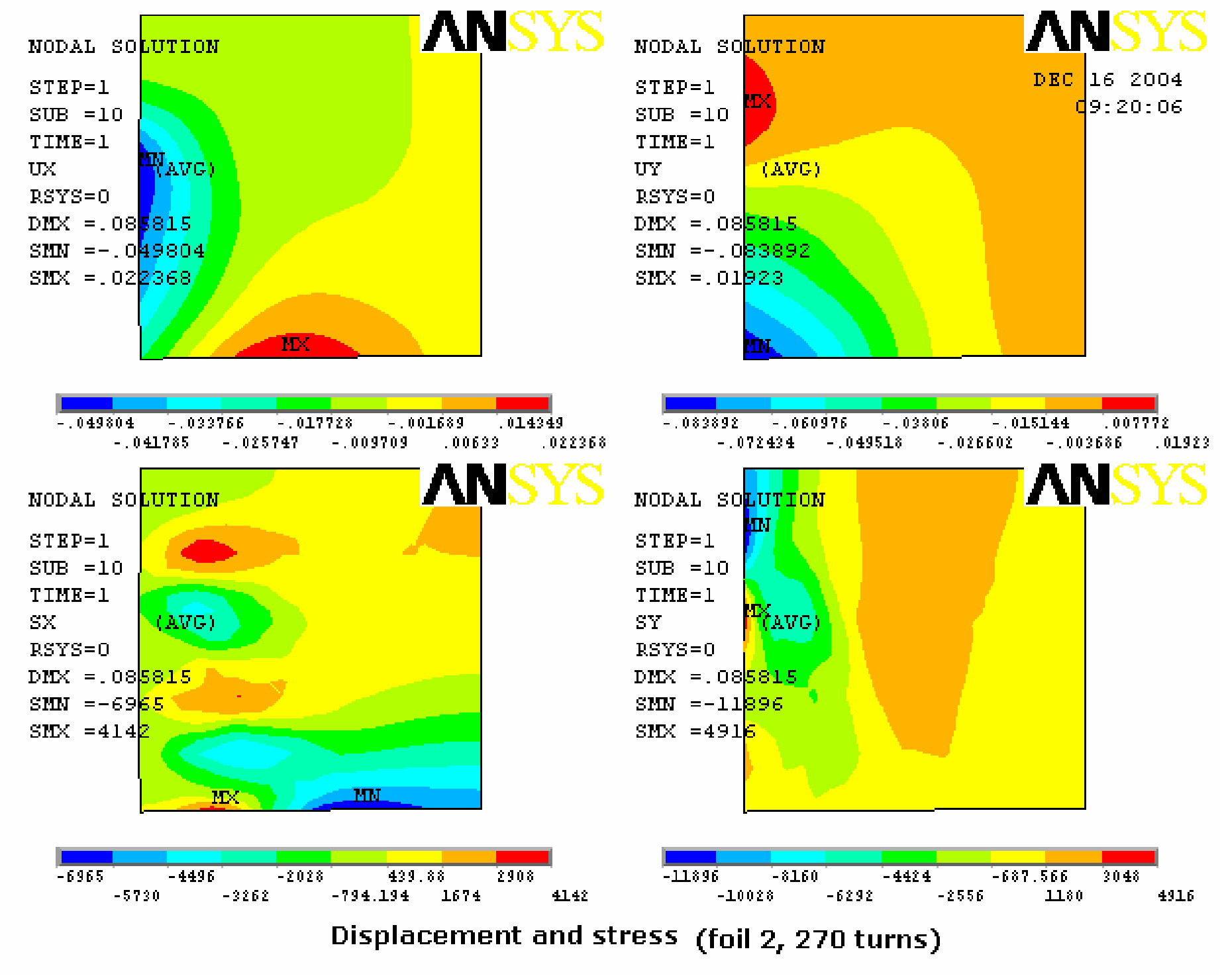

**Figure 16**

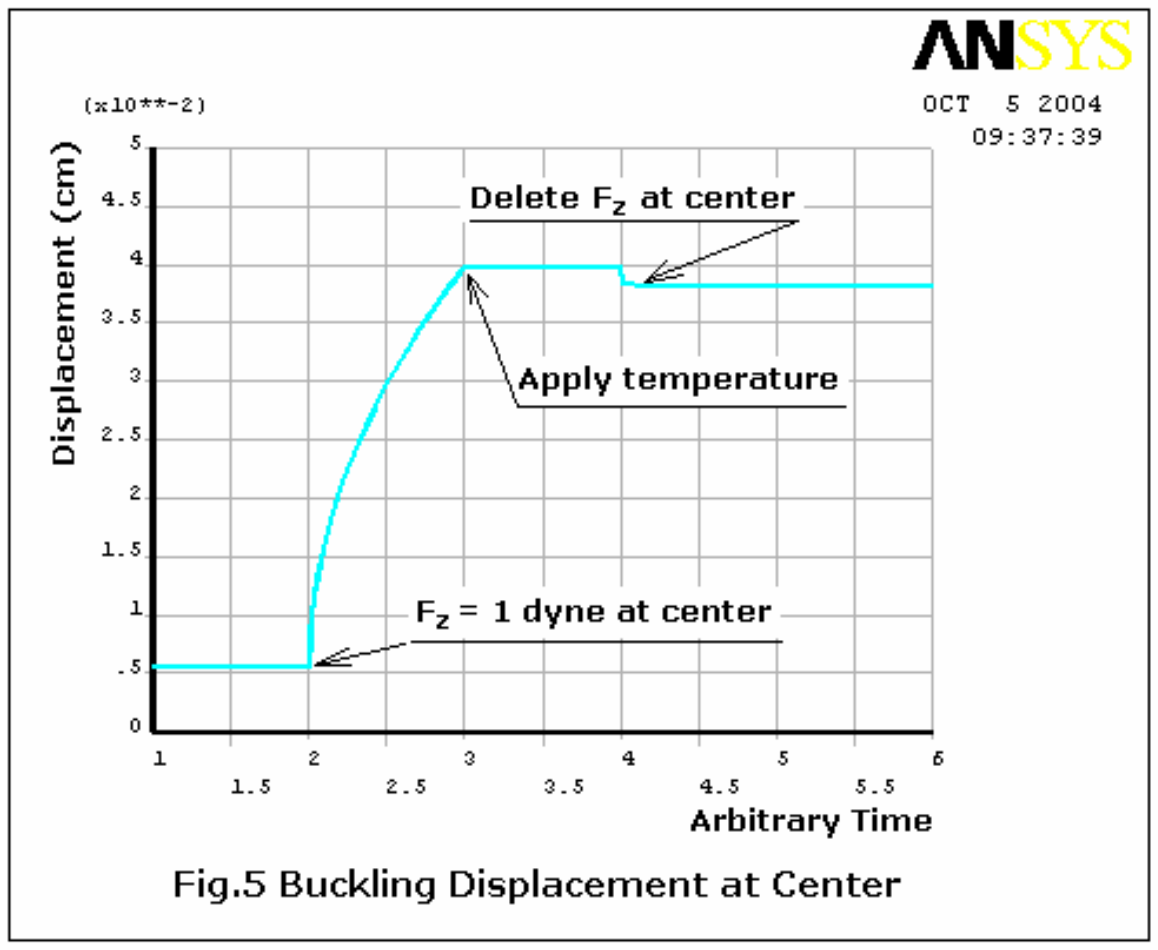

**Figure 17**

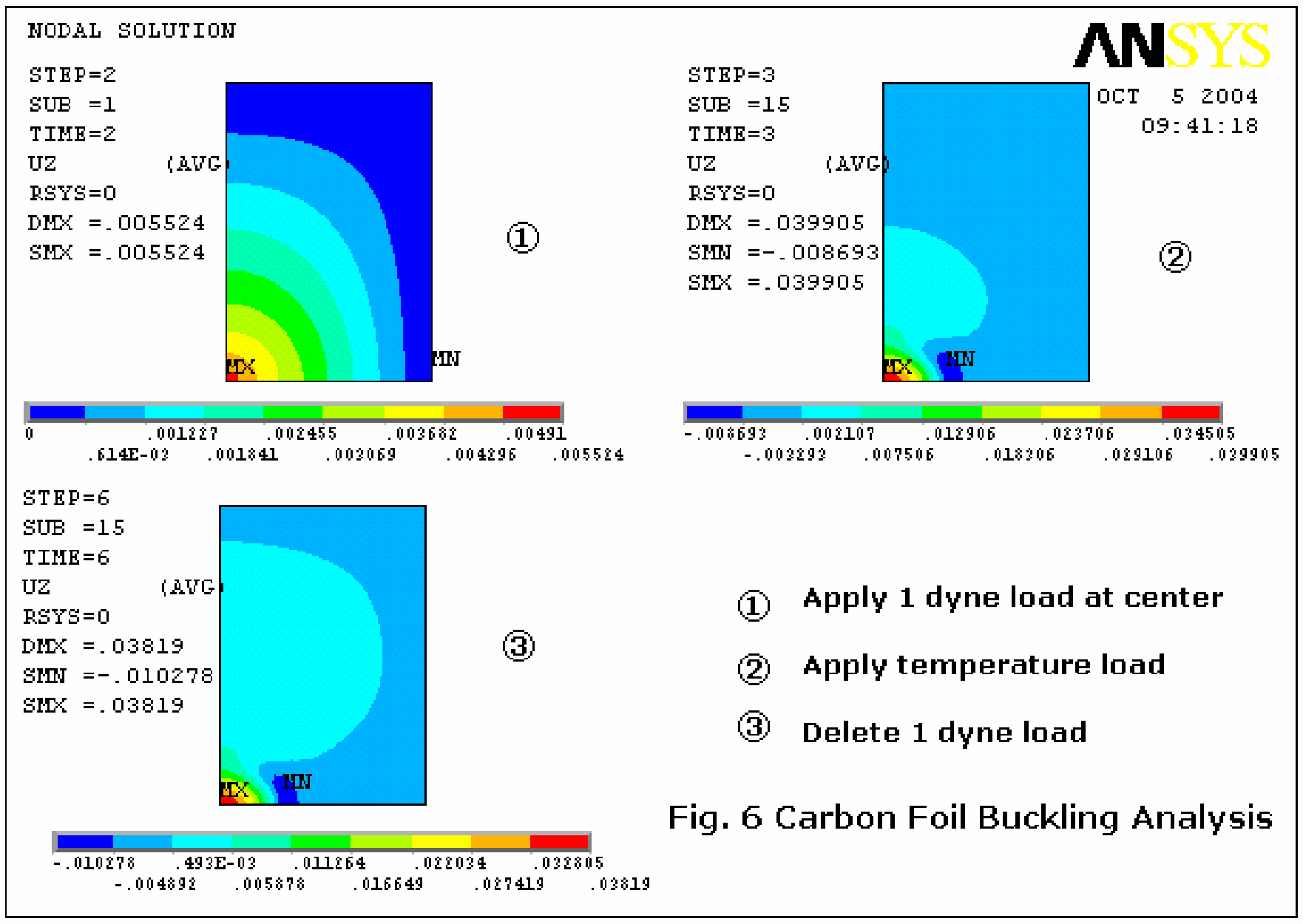

**Figure 18**

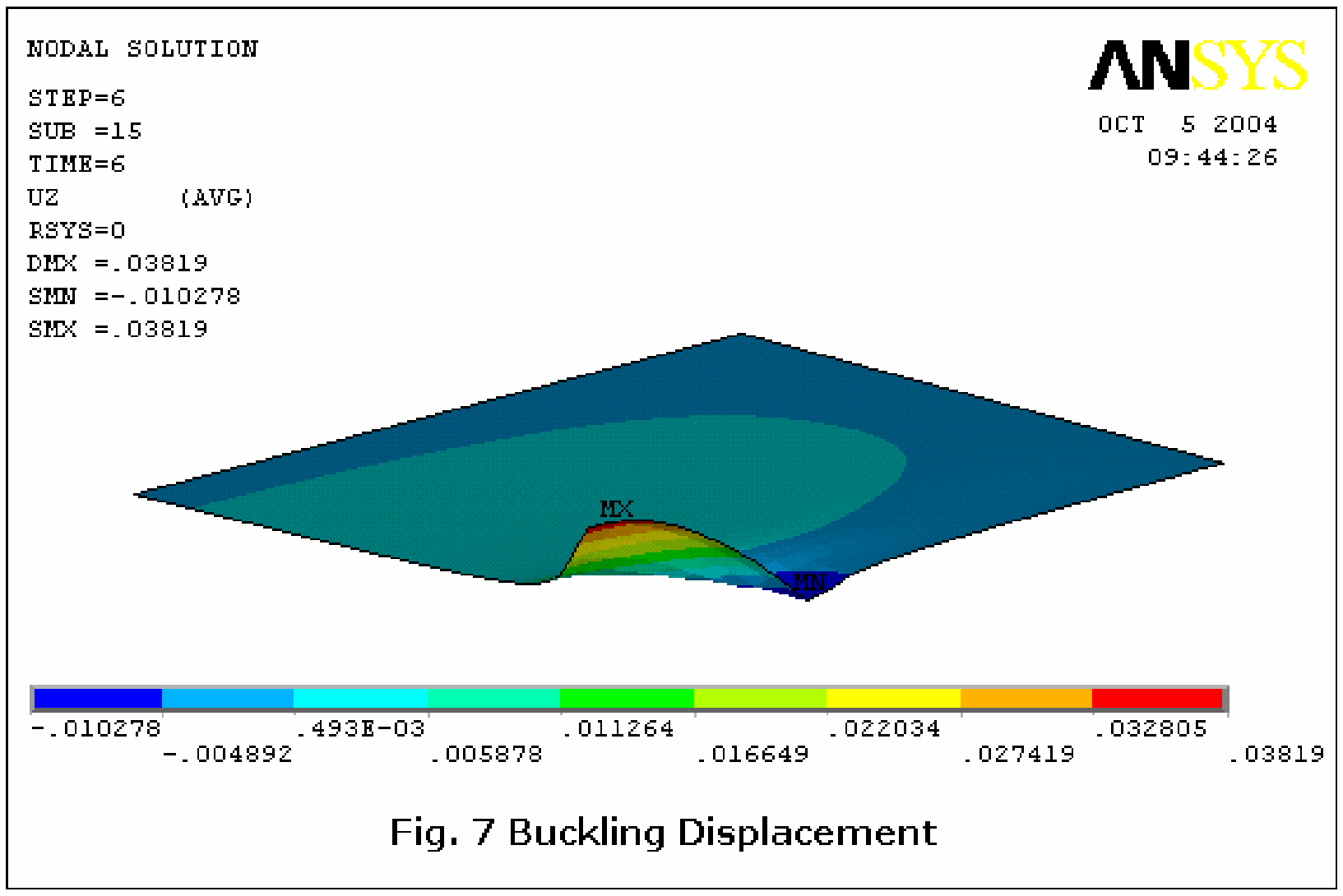

Fig. 7 Buckling Displacement

**Figure 19**

**Figure Captions:**

**Figure 1:** Left – the foil changer in the Fermilab Booster, right – an enlarged view.

**Figure 2:** Used carbon foils.

**Figure 3:** Total electron loss cross section for $H^-$ incident on H and He as a function of energy. (Ref. [2])

**Figure 4:** Total electron loss cross section for $H^-$ incident on N, O, Ar and Xe as a function of energy. (Ref. [3])

**Figure 5:** Measurement of $H^-$ stripping by carbon foil at 200 MeV. (Ref. [8])

**Figure 6:** Measurement of $H^-$ stripping by carbon foil at 800 MeV. (Ref. [7])

**Figure 7:** $H^0$ yield calculation using the cross sections in Table 1.

**Figure 8:** Instantaneous temperature rise for the 90-turn injection scheme. (MARS) Left – first foil, right – second foil. (unit: cm)

**Figure 9:** Instantaneous temperature rise for the 270-turn injection scheme. (MARS)

Left – first foil, right – second foil. (unit: cm)

**Figure 10:** Temperature history at the center of the foil. (ANSYS)

**Figure 11:** Temperature Distribution for the 90-turn injection scheme. (ANSYS)

Left – first foil, right – second foil.

**Figure 12:** Temperature distribution for the 270-turn injection scheme. (ANSYS)

Left – first foil, right – second foil.

**Figure 13:** Displacement and stress: Foil 1, 90-turn. (ANSYS)

**Figure 14:** Displacement and stress: Foil 2, 90-turn. (ANSYS)

**Figure 15:** Displacement and stress: Foil 1, 270-turn. (ANSYS)

**Figure 16:** Displacement and stress: Foil 2, 270-turn. (ANSYS)

**Figure 17:** Buckling displacement at foil center. (ANSYS)

**Figure 18:** Buckling distribution on the foil. (ANSYS)

**Figure 19:** 3-D plot of buckling displacement. (ANSYS)